\documentclass[doublespacing]{elsart}

\usepackage{graphicx}
\usepackage{amssymb}

\begin{document}

\begin{frontmatter}

\journal{SCES'2001: Version 1}
\title{ARPES Study of X-Point Band Overlaps in LaB$_6$ and SmB$_6$ -- 
        Contrast to SrB$_6$ and EuB$_6$}
\author[1]{S.-K. Mo}
\author[1]{G.-H. Gweon\corauthref{a}}
\corauth[a]{Corresponding Author. Phone: +1-734-647-9434, Fax:
  +1-734-763-9694, Email: gweon@umich.edu}
\author[2]{J.D. Denlinger}
\author[1]{H.-D. Kim}
\author[1]{J.W. Allen}
\author[3]{C.G. Olson}
\author[4]{H. H\"ochst}
\author[5]{J.L. Sarrao}
\author[6]{Z. Fisk}

\address[1]{Randall Laboratory of Physics, University of Michigan, Ann
  Arbor, MI 48109, USA}
\address[2]{Advanced Light Source, Lawrence Berkeley National Laboratory,
  Berkeley, CA 94720, USA}
\address[3]{Ames Laboratory, Iowa State University, Ames, Iowa 50011, USA}
\address[4]{Synchrotron Radiation Center, University of Wisconsin,
  Stoughton, WI 53589, USA}
\address[5]{Los Alamos National Laboratory, Los Alamos, NM 87545, USA}
\address[6]{National High Magnetic Field Laboratory, Florida State
  University, Tallahassee, FL 32316, USA} 

\begin{abstract}
  In  contrast to our  recent finding of  an X-point  band gap in divalent
  hexaborides,  we  report  here   that  angle   resolved  photoemission
  spectroscopy  (ARPES) data  shows  that the  gap  is  {\em  absent}  for
  trivalent LaB$_6$ and is  absent or nearly  so for mixed valent SmB$_6$.
  This finding demonstrates  a nontrivial evolution  of the band structure
  from divalent to trivalent hexaborides.
\end{abstract}


\end{frontmatter}


Rare earth hexaborides continue  to challenge condensed matter  physicists
to explain their   exotic properties,  e.g.,   the weak high   temperature
ferromagnetism   in lightly  doped   or nominally  stoichiometric divalent
hexaborides \cite{young}.   An  important issue is  the existence  or  the
absence of  band  overlap  at the X    point of the cubic  Brillouin  zone
appropriate   for these materials.  Such an   overlap is predicted by band
calculations  and would render divalent   hexaborides to be  semi-metallic
conductors, whereas they  would be insulators if there  is an X-point gap.
Recent angle resolved photoemission  spectroscopy (ARPES) data reported by
our group \cite{Denlinger00} show  a band gap of  $\approx$ 0.8 eV for the
divalent hexaborides SrB$_6$  and EuB$_6$.  We have  since  found the same
result    for     CaB$_6$   and    Ca$_{1+\delta}$B$_6$.    We   initially
\cite{Denlinger00}  interpreted  the gap   as  characteristic only of  the
surface region probed by ARPES.  Strong motivation to re-interpret the gap
as  characteristic  of  the bulk   was   provided by a  band   calculation
\cite{Tromp00} which includes a  GW self energy and  finds for  CaB$_6$ an
X-point gap of the size we observed.  By  complementary data from our bulk
sensitive  x-ray absorption and    emission  spectroscopy (XAS  and   XES)
\cite{Denlinger01}, we  have  now shown  conclusively  that a bulk X-point
band gap does indeed exist in these divalent  hexaborides.  In this paper,
we show that this gap is absent in  trivalent LaB$_6$ and either absent or
nearly so  in  mixed valent SmB$_6$.   This  finding demonstrates that the
band structure of the hexaborides  undergoes a  non-trivial change as  the
valency of the metal atom changes from divalent to trivalent.

Figures 1 and 2  show grey scale  intensity images of  energy vs.  crystal
momentum from   angle resolved  photoemission  spectroscopy  (ARPES)  data
obtained at the  Ames/Montana beam line, equipped  with a 50mm radius  VSW
spectrometer, of the Synchrotron Radiation  Center (SRC) at the University
of  Wisconsin.    The measurements were  done  on  cleaved  single crystal
surfaces  at   a temperature   of 20  K  and  in  a  vacuum   of $\approx$
4.0$\times$10$^{-11}$ Torr,  using 22 eV  photons.  Shown  in panel  B  of
Fig.\  1 is  a band calculation  for  LaB$_6$ \cite{Kubo89}.  The data are
presented   as a function   of   the momentum   value projected   onto the
$\Gamma$-X  line  of the Brillouin  zone.  Strictly  speaking, the k paths
corresponding to  the data are    arcs in  the $\Gamma$-X-M  plane,    not
necessarily close to  the $\Gamma$-X line.   However, extensive  data sets
show that these data are  nonetheless  representative of the states  along
the $\Gamma$-X line.   We tentatively attribute  this to a partial loss of
perpendicular  k-conservation   due    to   the  photo-electron   lifetime
\cite{hedin}.

The data for LaB$_6$ bear a strong resemblance to the theory of panel B\@.
We show the comparison  in panel A.  The band  $\alpha$ of La 5d character
agrees excellently with the data, and the band labeled ($\beta^\prime$) is
the theoretical  $\beta$ band of  B 2p character shifted  by 0.38 eV.  The
theoretical X point band overlap is 0.54 eV, and therefore we estimate the
experimental band overlap to be 0.16 eV.

In   the SmB$_6$ data  of Fig.\  2, both  the   $\alpha$-like band and the
$\beta$-like band are visible,  the latter more  strongly than the former.
Since the $\alpha$-like band is very weak in  the raw data (A), we enhance
its visibility in panel B by dividing the raw data by 11x11-point smoothed
data. Relative to  LaB$_6$, the Fermi energy lies  nearer to the bottom of
the $\alpha$ conduction band, as expected from  the reduced valency of the
Sm ion compared to the  La ion.  Another  difference  of the SmB$_6$  data
from the LaB$_6$ data is  the presence of  essentially k-independent Sm 5f
emission, faintly visible near $E_F$  in the data of  Fig.\ 2 and  clearly
observed in data taken  with higher photon energy  \cite{Denlingersces00}.
In Fig.\ 2,  we show as  lines  experimental band  dispersions obtained by
following the maxima of the image.  The position of the $\alpha$-like band
is especially uncertain   near  the X   point, due  to  the  weak emission
strength.  Our best estimate  is that the two  bands barely touch at the X
point, but a gap on the order of 0.1 eV is also possible.

Our findings here also agree with XES  and XAS data \cite{Denlinger01} and
show that there are non-trivial changes  in the band structure as divalent
hexaborides   evolve towards   trivalent  hexaborides.   Additionally,  an
anomalous deviation from Vegard's law  exists in the Ca$_{1-x}$La$_x$B$_6$
series.  Rather than  the monotonic increase  in lattice constant expected
from the larger La$^{3+}$ size, the  lattice parameter at first shrinks to
a minimum at $\approx$10\%  La doping before  increasing again with higher
La-doping \cite{Bianchi-tbp}.  As electrons are forced into the conduction
band   by  the alloying, the purely   single  particle contribution to the
ground state energy is reduced if the material can decrease its conduction
band energy by contracting,  up to the  point where the  gap is reduced to
zero and overlap begins.  It is interesting to speculate that the critical
La concentration in this  system  corresponds to the  band crossover.   We
note a tendency in both  LaB$_6$ and SmB$_6$ for  the overlap to ``stick''
near zero as  the chemical potential continues  to rise in  the conduction
band.

This work was supported  at  University of Michigan   by the US DOE  under
Contract No.    DE-FG-02-90ER45416  and by the US     NSF under Grant  No.
DMR-99-71611. The  Ames Lab is supported by  the US DOE under Contract No.
W-7405-ENG-82    and   the  SRC   is supported   by    the US   NSF  Grant
No. DMR-00-84402.



\begin{figure}
\centering
\includegraphics{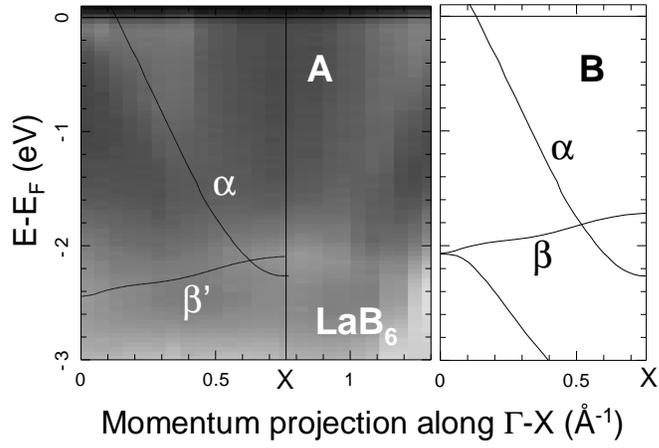}
\caption{ARPES data for LaB$_6$ (A) and theoretical band 
  calculation for LaB$_6$ (B) \cite{Kubo89}. For details, see text.}
\end{figure}

\begin{figure}
\centering
\includegraphics{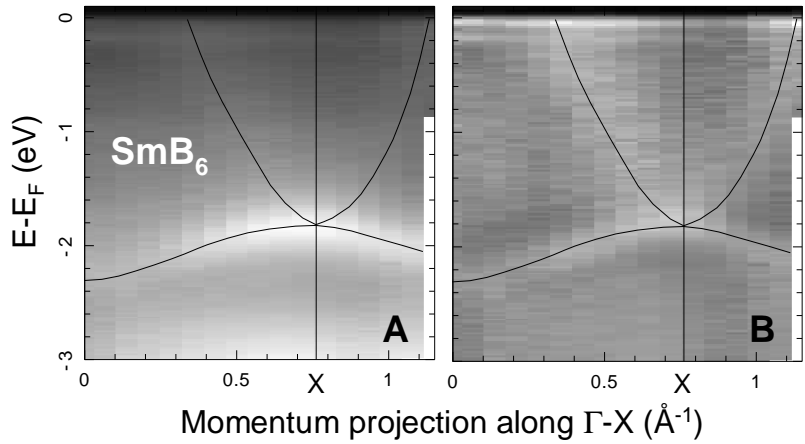}
\caption{ARPES data for SmB$_6$. For details, see text.}
\end{figure}


\begin{thebibliography}{06}

\bibitem{young}
D.~P. Young, {\it et~al.}, Nature 397 (1999) 412.
\bibitem{Denlinger00}
J.~D. Denlinger, {\it et~al.}, cond-mat/000922 (2000).
\bibitem{Tromp00}
H.J. Tromp, {\it et~al.}, cond-mat/0011109 (2000).
\bibitem{Denlinger01}
J.D. Denlinger, {\it et al}, to be published.
\bibitem{Kubo89}
Y.Kubo and S. Asano, Phys. Rev. B 39 (1989) 8822.
\bibitem{hedin}
W. Bardyszewski and L. Hedin, Physica Scripta 32 (1985) 439.
\bibitem{Denlingersces00}
J.~D. Denlinger, {\it et~al.}, Physica B 281 \& 282 (2000) 716.
\bibitem{Bianchi-tbp}
A.~D. Bianchi and Z. Fisk, unpublished.

\end{thebibliography}
\end{document}